\documentclass{epl}
\def\stackunder#1#2{\mathrel{\mathop{#2}\limits_{#1}}}
\newcommand{\mean}[1]{\left\langle#1\right\rangle}
\newcommand{\text}[1]{\hbox{#1}}
\newcommand{\sign}{\mathop{\rm sign}}
\newcommand{\erf}{\mathop{\rm erf}}
\newcommand{\e}{{\rm e}}
\newcommand{\euler}{\gamma_{\scriptscriptstyle E}}

\newcommand{\cc}{{c}}
\newcommand{\I}{{\rm I}}
\title{Partial survival and inelastic collapse for a randomly accelerated particle}
\shorttitle{Randomly accelerated particle}
\author{G. De Smedt\inst{1} \and C. Godr\`eche\inst{2} \and J.M. Luck\inst{1}}
\institute{
\inst{1}Service de Physique Th\'eorique, CEA Saclay,
91191 Gif-sur-Yvette cedex, France\\
\inst{2}Service de Physique de l'\'Etat Condens\'e, CEA Saclay,
91191 Gif-sur-Yvette cedex, France
}
\begin{document}
\maketitle
\begin{abstract}
We present an exact derivation of the survival probability
of a randomly accelerated particle
subject to partial absorption at the origin.
We determine 
the persistence exponent and the amplitude associated to the decay of
the survival probability at large times.
For the problem of inelastic reflection at the origin,
with coefficient of restitution $r$,
we give a new derivation of the condition for inelastic collapse,
$r<r_c=\e^{-\pi/\sqrt{3}}$,
and determine the persistence exponent exactly.
\end{abstract}

Consider a randomly accelerated particle,
obeying the stochastic equation of motion
\begin{equation}
\frac{d^{2}x_{t}}{dt^{2}}=\eta_{t},
\label{def1}
\end{equation}
where $\eta _{t}$ is Gaussian white noise with zero average,
$\left\langle\eta _{t}\right\rangle =0$, and correlator
$\left\langle\eta _{t}\eta
_{t^{\prime}}\right\rangle =\delta(t-t^{\prime})$.
This is the original Langevin equation with no damping force.
The joint random variables $(x_{t},v_{t})$ evolve in time as
\[
\frac{dv_{t}}{dt}=\eta_{t},\qquad\frac{dx_{t}}{dt}=v_{t},
\]
with initial conditions ($x_{0},v_{0}$).
Therefore
\[
v_{t}=v_{0}+W_{t},\qquad x_{t}=x_{0}+v_{0}t+\int_{0}^{t}du\,W_{u},
\]
where $W_{t}$ is the integral of the noise, or Brownian motion,
hence the process $x_{t}$ is  usually referred to
as the integral of Brownian motion.

The statistics of the times of first passage by the origin, 
and of related quantities, for a 
particle obeying~(\ref{def1}),
has been the subject of a long series of works, and is by now well understood
\cite{mckean,goldman,sinai,burk1,lachal}.

More recently, a number of studies have been devoted to survival
problems for a randomly accelerated particle,
with particular choices of the boundary conditions at the origin,
motivated by situations of physical interest.

In the partial survival problem,
the particle is absorbed with probability $1-p$ when crossing the origin,
and survives with probability $p$~\cite{majum}.
One is interested in the survival probability,
that is the probability for a particle obeying~(\ref{def1})
not to be absorbed at the origin up to time~$t$.
This reads
\[
G(t,p)=\sum_{n=0}^{\infty}p^{n}p_{n}(t),
\]
where $p_n(t)$
is the probability of the occurrence of 
$n$ zero crossings between $0$ and $t$.
At large times, the survival probability decays as
\begin{equation}
G(t,p)\stackunder{t\to\infty}{\approx}A(p)\,t^{-\phi(p)},
\label{jph}
\end{equation}
defining the persistence exponent $\phi(p)$~\cite{majum}
and the amplitude $A(p)$.
Very recently Burkhardt obtained the exact result~\cite{burk2}
\begin{equation}
\phi(p)=\frac{1}{4}-\frac{3}{2\pi}\arcsin\frac{p}{2},
\qquad\hbox{i.e.,}\quad p=2\sin\left[\frac{\pi}{6}(1-4\phi)\right].
\label{phi}
\end{equation}
He also obtained the persistence exponent $\Theta(r)$ for the inelastic
collapse problem, where the particle rebounds inelastically at the origin,
with coefficient of restitution $r<1$, i.e.,
the velocities $v_i$ just before and $v_f$ just after the collision obey
$v_{f}=-rv_{i}$~\cite{cornell}.
The result is \cite{burk2}
\begin{equation}
r^{2\Theta(r)}=2\sin\left[\frac{\pi}{6}\big(1-4\Theta(r)\big)\right].
\label{theta}
\end{equation}
Hence the two exponents are related by
\[
\Theta(r)=\phi\left(r^{2\Theta(r)}\right),
\]
as conjectured in ref.~\cite{swift}.

The purpose of this work
is to give new and simple derivations of these results,
relying on the statistics of the successive times of passage by the origin
of a randomly accelerated particle.
We shall also give an analytical prediction for the amplitude $A(p)$ of the 
power-law (\ref{jph}) \cite{burk3}.
We use the framework of ideas and results contained in
the classical work of McKean~\cite{mckean},
complemented by subsequent studies by Lachal~\cite{lachal}.

We begin by the partial survival problem.
Consider a particle starting from the origin with initial speed
$v_{0}$.
By definition,
\[
p_{n}(t)=\mathcal{P}(t_{n}<t<t_{n+1}),
\]
where $t_{n}$ is the time of occurrence of the $n$-th zero crossing,
with probability density $f_{t_{n}}(t)$.
Hence, in Laplace space,
\[
\hat{p}_{n}(s)=\frac{\hat{f}_{t_{n}}(s)-\hat{f}_{t_{n+1}}(s)}{s},
\]
where~\cite{lachal}
\[
\hat{f}_{t_{n}}(s)=\frac{2}{\pi}\int_{0}^{\infty}d\gamma\,\frac{\gamma
\sinh(\pi\gamma/2)}{\big(2\cosh(\pi\gamma/3)\big)^{n}}\frac{%
K_{i\gamma}\left(v_{0}\sqrt{8s}\right)}{v_{0}\sqrt{8s}},
\]
in which $K_{i\gamma}$ is the modified Bessel function.
After summation over $n\ge0$, we find
\begin{equation}
\hat{G}(s,p)=
\frac{1}{s}\left(1-(1-p)\frac{2}{\pi}\int_{0}^{\infty}
d\gamma\,\frac{\gamma\sinh(\pi\gamma/2)}{2\cosh(\pi\gamma/3)-p}\frac{%
K_{i\gamma}\left(v_{0}\sqrt{8s}\right)}{v_{0}\sqrt{8s}}\right).
\label{Ghat}
\end{equation}

We now have to extract the singular behavior of this expression
as $s\to0$, corresponding to the decay of $G(t,p)$ at large $t$.
To this end we use the integral representation
\[
K_{i\gamma}(x)\sinh\frac{\pi\gamma}{2}
=\int_{0}^{\infty}du\,\sin
(\gamma u)\sin(x\sinh u)\qquad(x>0),
\]
and the integral
\[
\int_0^\infty d\gamma\,\frac{\gamma\sin(\gamma u)}{2\cosh(\pi\gamma/3)-p}
=-\frac{3}{\sqrt{4-p^2}}\frac{d}{d u}\frac{\sinh\big(2(1-\phi)u\big)}
{\sinh 3u},
\]
where $\phi$ and $p$ are related through~(\ref{phi}).
After an integration by parts, (\ref{Ghat}) becomes
\[
\hat{G}(s,p)=\frac{I(0)-I(s)}{s\,I(0)},
\]
where
\[
I(s)=\int_0^\infty du\,\cosh u\frac{\sinh\big(2(1-\phi)u\big)}{\sinh 3u}
\cos\left(v_0\sqrt{8s}\sinh u\right),
\qquad I(0)=\frac{\pi\sqrt{4-p^2}}{6(1-p)}.
\]

The behavior of the difference $I(0)-I(s)$ for $s\to 0$
is dominated by large values of $u$.
At exponential order, setting $y=\exp(u)$, we obtain
\[
I(0)-I(s)\approx\frac12\int_0^\infty dy\,y^{-2\phi-1}
\left[1-\cos\left(v_0\sqrt{2s}\,y\right)\right]
=\frac{\Gamma(1-2\phi)}{4\phi}\,\cos\pi\phi\,\left(2v_0^2s\right)^\phi,
\]
leading to the result \cite{burk3}
\begin{equation}
G(t,p)\stackunder{t\to\infty}{\approx}
\frac{3}{2\sqrt{\pi}\,\phi\,\Gamma\left(\phi+\frac12\right)}
\,\frac{1-p}{\sqrt{4-p^2}}\left(\frac{v_0^2}{2t}\right)^\phi.
\label{Gtp}
\end{equation}

For $p=0$, this expression reads
\[
G(t,0)=p_0(t)\stackunder{t\to\infty}{\approx}
\frac{3}{\sqrt{\pi}\,\Gamma(\frac34)}
\left(\frac{v_0^2}{2t}\right)^{\frac14}.
\]
The persistence exponent of a free randomly accelerated particle,
$\phi=\frac14$~\cite{sinai,burk2} (see also~\cite{goldman}),
is thus recovered.
For $p=1$, the exponent $\phi=0$, and the right side of~(\ref{Gtp}) equals unity,
in agreement with the identity $G(t,1)=1$.

The result obtained, equation~(\ref{Gtp}), allows the determination of the
successive cumulants of the number $N_t$ 
of zero crossings between times $0$ and $t$.
Indeed, setting $p=\exp(z)$, 
the generating function of these cumulants is given by
$\ln G(t,p)=\ln\mean{\exp(z\,N_t)}$.
Expanding~(\ref{Gtp}) as a power series in $z$, we thus obtain,
$\euler$ being Euler's constant,
\begin{equation}
\mean{N_t}\stackunder{t\to\infty}{\approx}
\frac{\sqrt{3}}{2\pi}\left(\ln\frac{t}{2v_0^2}-\euler\right)+\frac16,\qquad
\mean{N_t^2}-\mean{N_t}^2\stackunder{t\to\infty}{\approx}
\frac{2\sqrt{3}}{3\pi}\left(\ln\frac{t}{2v_0^2}-\euler\right)+\frac{11}{72},
\label{jcu}
\end{equation}
 and so on.
The mean density of zero crossings in logarithmic time is thus equal to 
$\sqrt{3}/(2\pi)$~\cite{majum}.

We now turn to the inelastic collapse problem.
We consider again a particle starting from the origin
with initial speed $v_{0}$.

Following ref.~\cite{cornell}, we note that equation~(\ref{def1}) is invariant
under the change of scale:
$x\rightarrow r^{-3}x$, $t\rightarrow r^{-2}t$, which
implies the rescaling $v\rightarrow r^{-1}v$ of the velocity.
Hence,
performing this change of scale immediately after each collision compensates the
reduction of speed due to inelastic reflection.
One can therefore express
the elapsed time $t_\I$ in the inelastic problem in terms of the
elapsed time $t$ of a randomly accelerated particle, elastically reflected
when reaching the origin, as~\cite{cornell}
\[
t_\I=t_{1}+r^{2}(t_{2}-t_{1})+r^{4}(t_{3}-t_{2})+\cdots
+r^{2N_{t}}(t-t_{N_{t}})=\int_{0}^{t}du\,r^{2N_{u}},
\]
where $N_{t}$, the  number of crossings between  $0$ and $t$, and
$t_{n}$, the time of occurrence of the $n$-th crossing, pertain to the
elastic problem.
According to ref.~\cite{cornell}, the sum in the right side
converges if $r<r_{c}=\mathrm{e}^{-\pi /\sqrt{3}}\approx 0.163$, 
demonstrating the
inelastic collapse of the particle after the finite time
\begin{equation}
\cc=\lim_{t\rightarrow\infty}t_{\I}
=t_{1}+r^{2}(t_{2}-t_{1})+r^{4}(t_{3}-t_{2})+\cdots =\int_{0}^{\infty
}du\,r^{2N_{u}}.
\label{tc}
\end{equation}
As first observed in numerical simulations~\cite{swift}, 
the tail probability of the collapse time $\cc$ decays asymptotically as
\begin{equation}
\mathcal{P}(\cc>\tau)
\stackunder{\tau\rightarrow\infty}{\sim}\tau ^{-\Theta(r)},
\label{power}
\end{equation}
defining the persistence exponent $\Theta(r)$,
the analytical expression of which is given by (\ref{theta})~\cite{burk2}.

We now present new derivations of these results.

As shown by McKean~\cite{mckean},
a simple consequence of the scaling properties of Brownian motion,
and of the Markovian character of the couple ($x_{t},v_{t}$),
is the existence of the identities (in distribution)
\begin{eqnarray*}
t_{n}&=&t_{n-1}+u_{n-1}^{2}\,a,\\
u_{n} &=&u_{n-1}\,b,
\end{eqnarray*}
where $u_{n}=|v_{t_{n}}|$, $a=t_{1}/v_{0}^{2}$, $b=u_{1}/|v_{0}|$.
Hence
\begin{eqnarray*}
u_{n} &=&|v_{0}|\,b_{1}b_{2}\ldots b_{n},\\
t_{n}-t_{n-1} &=&v_{0}^{2}\,\left(b_{1}b_{2}\ldots b_{n-1}\right)^2\,a_n,
\end{eqnarray*}
in which the couples ($a_{1},b_{1}$), ($a_{2},b_{2}$), etc. are independent,
with common density~\cite{mckean}
\[
f_{a,b}(a,b)=\frac{b}{\pi\sqrt{3}\,a^{2}}\text{e}^{-\frac{2}{a}(1-b+b^{2})}%
\erf\sqrt{\frac{6b}{a}}.
\]
Integrating upon $a$ yields the marginal distribution~\cite{mckean,cornell}
\begin{equation}
f_{b}(b)=\frac{3}{2\pi}\frac{b^{\frac32}}{1+b^{3}}.
\label{marginal}
\end{equation}

The expression of $\cc$ given by~(\ref{tc}) can now be rewritten as
\begin{equation}
\cc=v_{0}^{2}\,\left(
a_{1}+r^{2}b_{1}^{2}a_{2}+r^{4}b_{1}^{2}b_{2}^{2}a_{3}+\cdots\right),
\label{tc2}
\end{equation}
hence (in distribution)
\begin{equation}
\cc=v_{0}^{2}\,a+r^{2}b^{2}\,\cc.
\label{kesten}
\end{equation}
This is the fundamental equation for the determination 
of the distribution of the collapse
time $\cc$.
The class of problems represented by eqs.~(\ref{tc2}) and (\ref{kesten})
has been investigated extensively
(for a review see e.g.~\cite{vervaat}).
In the present case, simple arguments lead to
the following results, starting from (\ref{kesten}).
(For a more rigorous treatment, see~\cite{kesten}.)

The probability density of the random variable $\cc$ 
obeys the integral equation
\begin{eqnarray}
f_{\cc}(\cc)
&=&\int da\,db\,f_{a,b}(a,b)\int du\,f_{\cc}(u)
\,\,\delta\left(\cc-v_0^2a-r^2b^2 u\right)
\nonumber
\\
&=&\int da\,db\,f_{a,b}(a,b)\,\frac{1}{r^2b^2}\,
f_{\cc}\left(\frac{\cc-v_0^2\,a}{r^2b^2}\right).
\label{eq_int}
\end{eqnarray}
Inserting in the last expression a power law of the form
$f_{\cc}(\cc)\sim \cc^{-(1+\Theta)}$ for $\cc\to\infty$,
the powers of $\cc$ on both sides cancel out,
and we are left with the condition
\begin{equation}
g(\Theta)=\mean{(r^2b^2)^\Theta}=1,
\label{jc}
\end{equation}
which determines the  exponent $\Theta(r)$.
Equation~(\ref{marginal}) yields
\[
\left\langle b^{s-1}\right\rangle =\frac{1}{2\cos(\pi s/3)},
\]
hence
\begin{equation}
g(\Theta)=\frac{r^{2\Theta}}
{2\sin\left[\frac{\pi}{6}\big(1-4\Theta\big)\right]},
\label{jf}
\end{equation}
so that~(\ref{jc}) is equivalent to~(\ref{theta}).

The function $g(\Theta)$ is convex, i.e.,
$g''(\Theta)>0$,
and  $g(0)=1$.
Therefore (\ref{jc}) leads to an acceptable solution for 
the exponent
$\Theta$ (i.e., a positive one), as long as
\begin{equation}
g'(0)=\left\langle\ln(r^2b^2)\right\rangle<0.
\label{cond1}
\end{equation}
If this condition is fulfilled, then the infinite sum~(\ref{tc2}) 
converges with probability $1$,
i.e., the collapse time $\cc$ is finite, 
which is the condition for inelastic collapse.
Equation~(\ref{jf}) yields $g'(0)=2(\ln r+\pi/\sqrt{3})$,
so that~(\ref{cond1}) is equivalent to
$r<r_c=\mathrm{e}^{-\pi/\sqrt{3}}$.

Let us remark that, in the derivation of (\ref{jc}), 
we implicitly assumed that the argument of $f_{\cc}$ in (\ref{eq_int})
was dominated by $\cc$.
This is justified provided that 
$\left\langle (v_0^2 a)^\Theta\right\rangle\equiv
\left\langle (t_1)^\Theta\right\rangle$ be finite,
which actually holds since $\Theta(r)<\frac{1}{4}$ for $r>0$,
as can be seen from (\ref{jf}).

Contrary to the partial survival problem~(see~(\ref{Gtp})),
determining  the amplitude of the power-law~(\ref{power}) for 
the distribution of $\cc$
is,  in general, a very difficult task~\cite{vervaat,derrida,calan}.

Besides persistence exponents, 
the limiting distributions of observables such as the occupation time
$\int_{0}^{t}du\,(1+\sign x_{u})/2$
of the process~(\ref{def1}), or the instant $t_{N}$ of
last visit of the particle to the origin, before time $t$,
are also of interest.
For the case of a  particle
leaving the origin with zero initial speed, Lachal~\cite{lachal2} 
found  
\[
F(x)=\lim_{t\to\infty}\mathcal{P}\left(\frac{t_N}{t}<x\right)=
\frac{3}{2\pi ^{2}}\int_{0}^{\infty}dz\,
\frac{\mathrm{e}^{2z}}{z}
\,K_{0}(4z\,x^{-\frac12})
\erf\sqrt{6z}.
\]
Using the integral representation
\[
K_0(x)=\int_{0}^{\infty}du\,\e^{-x\cosh u},
\]
a simple computation leads to the density
\[
f(x)=\frac{d}{dx}F(x)=
\left(\frac{3}{2}\right) ^{\frac{3%
}{2}}\frac{x^{-\frac{3}{4}}}{\pi ^{2}}\int_{0}^{\infty}du\frac{\cosh u}{%
(2\cosh u-x^{\frac12})\sqrt{x^{\frac12}+\cosh u}},
\]
which diverges as
$f(x)\approx A x^{-\frac34}$, with 
$A=3^{\frac32}\pi^{-\frac52}\big(\Gamma(\frac14)\big)^2/16
\approx 0.224$, as $x\to0$,
and takes the finite value $f(1)=\sqrt3/(2\pi)$, at $x=1$.
This value coincides with the mean density of zero crossings 
in logarithmic time, as expected.

Determining the limiting distribution of the occupation time
of the process~(\ref{def1}) remains an open problem.

In conclusion, the approach presented here brings,
on the problems of partial survival and inelastic collapse of a
randomly accelerated particle, a viewpoint complementary to that of
ref.~\cite{burk2}, based on Fokker-Planck equations.
\acknowledgments

We are indebted to A.J. Bray for having aroused our interest in this problem,
and to T.W. Burkhardt for stimulating discussions.

\end{document}